\documentclass[preprint]{aastex}
\usepackage{graphicx}
 \singlespace

\begin{document}

\title{Line Structure in the Spectrum of FU Orionis}
\author{P. P. Petrov}
\affil{ Crimean Astrophysical Observatory,\\
 p/o Nauchny, Crimea 98409, Ukraine}
\and
\author{G. H. Herbig}
\affil{Institute for Astronomy, University of Hawaii,\\
 2680 Woodlawn Drive, Honolulu, HI 96822 }

\begin{abstract}
 
New high-resolution spectra of FU Ori, obtained with the HIRES spectrograph
at the Keck I telescope in 2003--2006, make it possible to compare the optical
line profiles with those predicted by the self-luminous accretion disk
model. A dependence of line width on excitation potential and on wavelength,
expected for a Keplerian disk, is  definitely not present in the optical
region, nor is the line duplicity due to velocity splitting.
The absorption lines observed in the optical
region of FU Ori must originate in or near the central object, and here
their profiles are shown to be those expected of a rigidly rotating object.
They can be fitted by a rapidly rotating ($v$ sin $i$ = 70 km s$^{-1}$)
high-luminosity G-type star having a large dark polar spot, with axis
inclined toward the line of sight.  Over these years, the radial velocity of
FU Ori has remained constant to within $\pm$0.3 km s$^{-1}$, so there is
no indication that the star is a spectroscopic binary.  These results
apply to the optical region ($\lambda< 8800$ \AA); more distant, cooler
regions of the disk contribute in the infrared.

\end{abstract}

\keywords{stars: evolution --- stars: individual (FU Orionis) --- stars: 
pre--main-sequence --- stars: variables: other }

\section{Introduction}

	FU Ori is the prototype of a small group of pre-main-sequence
stars, the so-called FUors.  As explained in the Introduction to
\citet{herb03}, the first members were optically detectable
and so could be studied in some detail.  They were distinguished
by flareups of several magnitudes followed either by near-constancy
or slow decline; by broad-lined absorption spectra resembling F- to K-type
supergiants in the optical region; by powerful outflowing winds shown
by P Cyg structure at H$\alpha$ and the \ion{Na}{1} D lines; and by
high \ion{Li}{1} abundances (an indication of youth). All the known FUors
are located in star-forming regions.  These characteristics
defined the FU Ori class; their observed properties have been reviewed
in detail by \citet{herb77} and by \citet{hart96}.

 	In recent years, some authors have chosen to take a broader
and more inclusive view of  the FUor phenomenon. For example, the list
of \citet{abr04} contains some stars that are certainly
not FUors
in the original sense.  We see no advantage in lumping such incompletely 
characterized  objects in with those about which we know a great deal,
such as FU Ori and V1057 Cyg. So we prefer to call them ``FU Ori-like" 
(following \citealt{gre08}) until their status can be clarified.  This paper
deals with FU Ori.  

	One of best-documented FUors is V1057 Cyg, which before its flare-up
by about 6 mag in 1971 was a faint variable star with an emission-line
spectrum like that of a T Tauri star. After the flare-up, it appeared as
a high-luminosity early-type star with only H$\alpha$ in emission. It has
since been in slow decline, and is now (2008) only about 1 mag (in B)
above its pre-outburst brightness.

	FU Ori itself increased in brightness from $m_{\rm pg} = 16$ to 10 in
 1937--1939, and since that time has remained near maximum, except for a
 slow decrease in optical brightness of about 0.15 mag per 10 years
 \citep{cl05}.  Before the flare-up, FU Ori was an irregular variable of
 unknown spectral type. Optically, it is now  classified as G0, about
luminosity class Ib. The spectral energy distribution shows an IR excess,
typical for a pre--main-sequence star with circumstellar disk and/or dust shell.
There is also excess flux in the UV region ($\lambda$ $<$ 2600 \AA) indicating
 the presence of a hot continuum of $T_{\rm eff} \approx  9000$ K \citep{krav07}.

	FU Ori has a faint red companion, probably a pre--main-sequence star, at a projected separation of 0\farcs5 = 250 AU \citep{wang04}.
It is unknown whether the presence of this star has a connection to the
 outbursts.

	A widely known concept is that FUors are solar mass pre--main-sequence
stars that have increased dramatically in brightness due to a sudden episode of
enhanced accretion triggered by some kind of disk instability. The rate of
accretion has been estimated as $2 \times 10^{-4} M_\odot$ yr$^{-1}$
by \citet[hereafter PKHN96]{pop96} by fitting an accretion disk model to
the observed spectral energy distribution.  That is three orders of magnitude
larger than in a classical T Tauri star. The optical spectra of FUors contain
broad metallic lines that, according to \citet[hereafter KHH88]{ken88},
originate in the atmosphere of the inner accretion disk at radii of 8--80
$R_\odot$.

	There is general agreement that the IR excess seen in the SED of FU
Ori originates in an extended envelope---presumably a disk---whose outline
is seen in the $K$-band interferometric image of \cite{mal05}. In the
model of Hartmann \& Kenyon, this structure extends inward to the central
star and overwhelms it: ``during outburst, the disk outshines the central star
by factors of 100--1000" \citep{hart96}.  Thus the optical spectrum of FU Ori
is thought to be formed in an atmosphere atop the surface of this
optically thick disk, and the contribution of the central star is 
small or negligible.

	Because of the plausibility of this hypothesis, and the many subsequent
papers that built upon it,  it largely escaped the critical
scrutiny that usually follows upon such an interesting new proposal. 
A simple yet critical test of the self-luminous accretion disk model can
be provided by the higher-resolution optical spectra that have now become
available.  These make it possible to approach the key question: whether
the {\it optical} line profiles are in fact those of a Keplerian disk or not.

	One of the most explicit predictions of that FUor disk model is that
line width (or $v$ sin $i$) should depend on wavelength and excitation
potential, because the warm central regions of the disk rotate faster than
the cooler regions farther out. Observational evidence for such relationships
in the optical spectra of FU Ori and V1057 Cyg were reported by \citet{welt92},
but not confirmed later by spectroscopic data of higher quality
and resolution (\citealt{herb03}, hereafter HPD03).  This paper deals with that
issue in more detail.

	Another prediction of the model was that FUor absorption lines should
appear double because of Keplerian splitting.  No such splitting is
observed in the high-resolution spectra of FU Ori described in this
paper.  In V1057 Cyg, some
(but not all) absorption lines appeared to be double, but as the star
faded, it became clear that the duplicity  was caused by the emergence of
emission cores in the line centers that are produced in a hot region (which was
called a ``chromosphere" by HPD03 for lack of a better word).

\section{Observations}

  	This paper reports an analysis of absorption line profiles of
FU Ori on new spectra obtained beween 2003 and 2006 with the HIRES
spectrograph on the Keck I telescope on Mauna Kea.\footnote{The W. M. Keck
Observatory is operated as a scientific partnership among the California
Institute of Technology, the Univerity of California, and the National
Aeonautics and Space Administration.  The Observatory was made possible by
the generous financial support of the W. M. Keck Foundation.}  The dates
of those observations are given in Table 1.  The 2003 spectrograms covered
the region 4350--6750 \AA , while the remaining four covered 4750--8690 \AA,
both with interorder gaps.  The nominal resolution was 48,000.  The slit
width projected on the sky was $0\farcs86 \times 7\arcsec$.  The
spectra were reduced by conventional IRAF procedures.  The velocity
zero-point of each exposure was checked by measures of atmospheric lines.

\section{Photospheric Line Profiles}

	The new spectra of FU~Ori appear similar to those
of 1995--2000, described in detail by HPD03, and to the single
observation of 1987 discussed by \citet{pet92}.   H$\alpha$ has a P Cyg profile
with a strong blueshifted absorption extending to about --400 km s$^{-1}$
that is produced by an outflowing wind.  A weaker longward emission component
is also present.  Both absorption and emission components vary from night
to night. The wind absorption is always strong, while the emission component
may disappear completely on some occasions (see \citealt{dan00} and Figs.\
26 and 32 in HPD03).

	On account of the high resolution and quality of the HIRES
spectra, the distinctive profiles of photospheric lines are quite apparent. A
sample of both strong and weak metallic lines is shown in Figure 1.  The weaker
lines have symmetric wings, while the stronger have either wider
shortward wings (e.g., \ion{Ca}{1} 6439) or a deep blueshifted absorption
component (e.g., \ion{Mg}{1} 5183, \ion{Fe}{2} 5018).  The stronger lines
are formed in  outer layers of the atmosphere; such profiles were described
by \citet{cal93} in terms of a disk wind model.

	The weaker lines have a ``boxy" shape, with steep wings and flat
bottoms (Fig. 1, right panel).  Fine structure can be seen at their bottoms, 
and sometimes narrow minima appear at one or both edges. Such profiles
persist on a timescale of months and years. 
Figure 2 shows a sequence of six spectra of the 7530--7590\,\AA\ region, 
which contains a few outstanding metal lines. A spectrum of $\beta$~Aqr 
(G0 Ib) taken with the same instrument is shown for comparison. 
In FU Ori, a slight profile asymmetry that changes from one spectrum to
another is present: either the shortward or longward side of the profile
becomes deeper. This variable asymmetry of photospheric lines was noticed
earlier in the SOFIN spectra of FU Ori (HPD03), where a period of 3.5 days
was suspected.  The time spacing of the present HIRES observations is so
large that although variations can be detected, the periodicity cannot
be verified.

	Here we concentrate on the profiles of weak photospheric lines,
which are  not affected by wind/shell absorptions and thus may carry
information about the structure of photospheric layers.  The equivalent
widths (EW) of \ion{Fe}{1} and other photospheric lines in FU Ori are about the
same as in a G-type supergiant, but due to their large widths, those lines
are very shallow.  The weakest lines that we could measure on the HIRES
spectra have EWs of approximately 0.05 \AA\ and central depths of about 0.015 
of the continuum.

	In most previous spectroscopic studies of FUors, cross-correlation
was used to improve the signal-to-noise ratio and to derive an ``average"
line profile over a certain spectral interval, while in these HIRES spectra
we are able to measure the profile of each line individually. 

	 Because of the large line widths, blending is a severe problem
especially in the blue, so minimally blended lines were selected.  Furthermore,
some lines are accompanied shortward by wind (or shell) components, so
only the longward half of those profiles could safely be considered to be
purely atmospheric.  Table 2 is a list of 59 lines selected as adequately
free of blends or wind structure. The symbol $r$ in the last column
means that only the longward half of that profile is blend-free. The line
parameters in Table 2 are taken from the Vienna Atomic Line Database
 \citep{kup99}.

	As a line width parameter, we selected the half-width at half-depth
(HWHD) of the longward side of the profile, as illustrated in Figure 3.
The standard deviation of one line measurement is 1.5--2.0 km\,s$^{-1}$.
The value of HWHD, averaged over all the lines of Table 2, is given for
each observation in Table 1, where $\sigma$ in the fourth column is the
standard deviation of the average in km s$^{-1}$.  The average HWHD
 is not the same on all six dates.  The overall average
is 61.88 $\pm$ 0.19 km\,s$^{-1}$, so the variations of HWHD, if real, 
do not exceed the $\sigma$ in the fourth column of Table 1.
Since HWHD is determined from the red wing only, any change in
the radial velocity of FU Ori would cause a change in HWHD as well.
Therefore, the constancy of HWHD within $\pm$0.19 km\,s$^{-1}$ 
is compatible with constancy of the radial velocity of FU Ori at that
level.

	 Radial velocity shifts can also be obtained by cross-correlating 
each of the six spectra of FU Ori with their average. This procedure was
done for the spectral intervals: 5995--6030, 6570--6620, and 7705--7765 \AA;
the results are listed as $\Delta$RV in Table 1.
These radial velocity shifts between spectra do not exceed $\pm$0.6
 km\,s$^{-1}$; the standard deviation is 0.33 km s$^{-1}$.  The accuracy of
this method is lower than that described
above because slight changes in profile asymmetry may partly contribute to the 
apparent velocity shifts (HPD03). It is safe to conclude that the radial
velocity of FU Ori derived from these six HIRES spectra was constant within
$\pm$0.3 km s$^{-1}$ over this time interval. 

\section{Comparison to Models}

	These new HIRES spectra of FU Ori make it possible to compare 
observations with the spectra predicted by accretion disk theory. 
In this section we first test whether line width depends on wavelength and
excitation potential (EP). Then we compare the observed line {\it profiles} 
to those calculated for two different models: (1) a differentially rotating 
Keplerian disk, and (2) a rigidly rotating spherical star with a dark polar
 spot.

\subsection{Disk Model}.

	The first models for FU Ori and V1057 Cyg, by KHH88, were thin
accretion disks
with temperature $T_{\rm eff}$ decreasing with distance $r$ from the disk center 
as $T_{\rm eff} \propto r^{-3/4}$ and Keplerian rotational velocity 
$v_{\rm rot} \propto r^{-1/2}$.  In PKHN96, the model was improved by including 
a self-consistent treatment of the boundary layer region.

	During the 20 years between the middle of 1980s, when the
accretion disk model (KHH88) was put forward, and the beginning of our
present HIRES observations in 2003, the brightness of FU Ori has dropped
by only about 0.3 mag in $B$, $V$, and R \citep{cl05}, so we assume that the
distribution of energy in the optical region did not change significantly,
and so we begin by comparing the KHH88 model with our observations. That
model predicts a specific dependence of line width on wavelength and excitation
potential (EP), a prediction that has in many subsequent papers on FUors
been cited as observational support for the theory.

	According to the KHH88 prescription, the disk can be considered
as a sum of
annuli, each of radius $r$, temperature $T_{\rm eff}(r)$, and projected rotational 
velocity $v\,\sin i$, also a function of $r$; $i$ is inclination of the
disk rotational axis to the line of sight.
We use the same parameters of the model as in KHH88: 
radius of the central star 5.5 $R_\odot$, inner radius of the disk
 $r_{\rm in} =1.46\ R_{*}$, peak temperature $T$ = 7200 K at $r_{\rm in}$,
 $v\,\sin i$ = 93 km\,s$^{-1}$ at $r$ = 1\, $R_*$; $v\,\sin i$ is the
 main parameter that determines spectral line width. 

	The intensity spectrum of each annulus can be represented by a
synthetic spectrum emerging from the stellar atmosphere 
at a certain angle $\theta$ between the line of sight and normal to the
stellar surface. These synthetic spectra were calculated using the code
by Berdyugina (1991) and Kurucz's stellar atmosphere models \citep{kur93}.
We chose $\theta = 45\arcdeg$ to make the spectra more appropriate for
modeling an inclined disk.  However, this parameter is not critical for
line profiles in the integrated spectrum of the disk, as long as $\theta
 < 60\arcdeg$. The synthetic spectra were convolved with the rotational
 function (defined in KHH88) of the corresponding annulus in the disk, and
convolved with a Gaussian profile corresponding to the spectral 
resolution of HIRES.

	In the inner 18 annuli, within $r$ = 6 $R_*$, the temperature
 decreases gradually from 7200 to 3500 K. Within this range, the following
 discrete set of $T_{\rm eff}$ was used for calculation of the synthetic
 spectra: 7200, 6500, 5750, 5000, 4500, 3750, 3500 K, with log $g=2$,
 and $v_{\rm micro}$ = 2 km\,s$^{-1}$.  This is about the same as the set
 of spectral templates from F2 to M1 types used in KHH88. Rotational
 velocity was calculated
 explicitly for each annulus as a function of its radius.
For the outer 5--8 annuli with $T_{\rm eff} < 3500$ K ($r$ = 6--13 $R_{*}$), 
high-resolution spectra of late-type stars were obtained 
from the ESO/UVES spectral library: HD 95950 (M2 Ib) and HD 34055 (M6 V).   

	The contribution from each annulus to the integrated spectrum 
(as defined by its temperature and surface area) was taken from the 
KHH88 model. Spectra from each annulus with its contribution factor 
were added up to produce the integrated spectrum of the disk.
The outer annuli with T$_{\rm eff}$ $<$ 3500 K contribute only a few percent
to the optical spectrum of the disk. Moreover, the atomic lines discussed
in this paper are formed within the inner regions of the disk, which
are represented
 by the Kurucz model spectra. Nevertheless, we attempt to exclude from
the  analysis those atomic lines that fall near the strongest breaks
in the continuum caused by molecular bands in the M-type spectra,
which might distort the local continuum level in the integrated spectrum. 

	The only difference between this procedure and that described in KHH88
is that we used synthetic spectra of stellar atmospheres, while in KHH88 
observed spectra of template stars were used. 
This may be important when fine structure of photospheric lines is
 considered. First, we examine whether line width is a function of EP and
 wavelength in both observed and calculated spectra.

	Table 3 is a list of lines measured in the disk model spectrum.
It is shorter than that in Table 2 because of more severe
blending in the predicted spectrum of the disk. The error in the measurement
of HWHD for one line is about the same ($\pm$2.0 km s$^{-1}$) as in
HIRES spectra.  Although model spectra are noiseless, there is always some
uncertainty in locating the local continuum level.  Figure 4 shows line width
vs.\ EP and vs.\ wavelength for the HIRES spectrum of highest quality
(2004 November 21) as filled circles, and for the model 
spectrum of the disk as open circles.  HWHD vs. EP (filled circles in the
left panel) show a correlation coefficient (CC) of 0.18, which with 59 data
points corresponds to a false alarm probability (FAP) of 20\%; i.e., there is
no correlation of HWHD with EP.  In any case, the line slope is opposite
to that expected from the disk model.

	On the right panel, HWHD vs.\ wavelength, the filled circles show
a weak correlation (CC = 0.34) in the sense of line width increasing
toward longer wavelength,  but again, that is in the opposite sense to that
predicted by the disk model.  As expected, the disk model (open circles)
predicts a clear dependence of line width on wavelength.
(The larger scatter of points in the model plot is caused by
the dependence of line width on both of these parameters.) 

	So, in contrast to the model prediction, the observed HWHDs in
FU Ori show no dependence on either EP or wavelength.  This result was
checked for each of the six spectra of FU Ori (see Fig. 5). It confirms
the conclusion of our previous paper (HPD03), but the higher quality of
the new Keck spectra makes the result more firm.

	The disk model also predicts a doubling of photospheric absorption
lines, which is not evident in these spectra of FU Ori, where the line profiles
 have rather flat bottoms.  Nevertheless, when the HIRES spectra of FU Ori
 are cross-correlated with the spectrum of a G0 Ib template ($\beta$ Aqr),
 the cross-correlation function is slightly doubled, similar to
 that seen in the earlier observations of 1995--2000  (HPD03), and in
the single spectrogram of 1987 \citep{pet92}.  This is due to the narrow
 absorption dips flanking the flat bottom of the line
profile; this effect is discussed in an alternative model below.
Line doubling may be also caused by the presence of emission cores in
the line centers, an effect that actually was observed in V1057 Cyg after
1995, when that star dropped in brightness and some emission cores rose
above the continuum level \citep{pet98}.  Such an emission core in
\ion{Li}{1} $\lambda$6707 may be present in some of our spectra of FU Ori.

	In our discussion \citep{pet92} of a spectrogram of FU Ori obtained
in 1987 January (at a resolution comparable to HIRES), we noted that although
not striking, a doubling of metallic lines was detectable at that time
by cross-correlation against a standard supergiant.  In the 
case of the HIRES spectra discussed here, although variable structure is
apparent, no duplicity is evident on cross-correlation.  The earlier
 SOFIN spectra showed that the line profiles change from night to night
(Fig. 29 in HPD03), so there is no evidence of a secular change in the
 line structure.

	The new HIRES spectra were also compared to those predicted by
the boundary layer model of PKHN96. In that model for FU Ori, the boundary
layer is an annulus extending from the stellar surface to about 2.2 $R_{*}$.
The expected rotational profile differs from the Keplerian in such a way that
 maximum rotational velocity of 47 km\,s$^{-1}$ is at 2.2 $R_{*}$,
while closer to the stellar surface the rotation decreases to 33 km\,s$^{-1}$.
Within the boundary layer, the temperature ranges from 8200 to 5400 K. Outside
the boundary layer the disk parameters approach those of the classical
Keplerian thin disk. Thus, the optical spectrum is formed mostly within the
boundary layer, and all spectral lines with EP from 2 to 8 eV have about the
same width. Only lines of 0--1 eV, formed in the distant regions of the disk,
are a few km\, s$^{-1}$ narrower. The dependence of line width on EP for the
PKHN96 model is shown by the dashed line in Figure 4 (left). It is much
lower than the observed curve because of the low rotational velocity in the
boundary layer, as specified by PKHN96 for FU Ori. The line at 6170 \AA\
used by PKHN96 to fit their model to observations is not a good indicator of
line width because it is a blend of three lines (\ion{Ca}{1} 6169.04, 6169.56 and
\ion{Fe}{1} 6170.51) that are of comparable strength at $T_{\rm eff}$ = 5000--6000 K.
The blend cannot be resolved because the width of each component is
$\approx$2.5 \AA, due to rotational broadening.

	We conclude that the observed line widths of FU Ori do not follow
 the relations predicted by the Keplerian disk model.  We remark also that
 the narrower widths of low excitation lines of the PKHN96 model would have
 been obvious on these HIRES spectra, if present.

	Still more information can be retrieved from the FU Ori line profiles. 
In Figure 6 observed line profiles  are compared
with those of the Keplerian accretion disk model. The main differences between
observed and modeled profiles---dependence of the model line width on EP,
and the line doubling---can be seen even in this small spectral region.
 When the PKHN96 model is used, the match is worse, because those modeled
lines are narrower and deeper than observed and have more prominent doubled
structure.

	The wings of lines in FU Ori are steeper than in the classical case
of a rotating spherical star having reasonable limb darkening. The Keplerian
disk also produces a rather smooth profile. A ring or torus (like the boundary
 layer) might produce lines with steep wings, but it would also 
produce a prominent hump in the line center. The flat bottom and steep
 wings that are actually observed are typical of another kind of object:
 a fast-rotating star with dark polar spot.  Therefore we now examine a
 simple ``star-spot" model (as was first applied to FU Ori by \citealt{herb89}).

\subsection {``Star-spot" Model}

	A dark spot on the surface of a rotating star produces a bump in the
 otherwise smooth rotationally broadened profile of an absorption line. As
the star
 rotates, the spot passes across the visible hemisphere, and the bump travels
 across the line profile. This effect is utilized in the Doppler imaging
 technique. If a large dark spot is located at the pole 
of rotation, line profiles remain disturbed throughout the period. 
Figure 7 shows a star with a polar spot and illustrates the consequence: a
 flat bottom in the line profile. Figure 8 shows how the
 line profile (of \ion{Ni}{1} 7555.59 \AA, as an example) responds to
 different model parameters. (There are similar spots on both poles in
 this model, but only one is visible.) Only two free parameters
 are required to fit the model to the observed profiles:
 the inclination of the rotational axis ($i$) and the spot radius ($R_{\rm spot}$).
The value of $v$\, sin $i$ = 70 km\, s$^{-1}$ is fixed by the observed width
 of lines at continuum level.

 An upper limit to the spot temperature can
 be set by the absence of TiO bands and other M-type features in the optical
region (see Fig. 9).  If the spot were hotter, the TiO band would be
apparent in the stellar spectrum.
 With $T_{\rm spot} < 3300$ K, the ratio of spot to photosphere brightnesses
 $B_{\rm spot}/B_{\rm phot}$ would be less than 0.17 in the red. With the spot area
occupying 30\% of the visible hemisphere in projection, the contribution
of the spot to the total spectrum would be less than 5\%. 

	The model spectrum of the spotted star was calculated using the
 synthetic intensity spectrum with $T_{\rm eff} = 5000$ K, log $g$ = 2,
 $v_{\rm micro}$ = 2 km\,s$^{-1}$ for the photosphere, and $T_{\rm eff} = 3500$ K
 for the spot.  Those spectra were convolved with the instrumental profile
 and integrated over the corresponding areas of the visible stellar
 hemisphere, with the limb darkening coefficient 0.7, inclination $i$,
 and $v$  = 70/sin $i$ km\, s$^{-1}$.

	In Figure 10 the observed line profiles are compared to those
 of the ``star-spot" model that results from this
 calculation.  The best match was achieved with
 $R_{\rm spot} = 40^\circ$ and $i = 37^\circ$ (there is 
 freedom of a few degrees in the choice of those parameters). Such a spot
 occupies 12\% of the global stellar surface (about 30\% in projection on
the visible stellar hemisphere).  Note the narrow absorption dips at both
 sides of the flat bottom, which appear at a certain inclination and spot
 radius. These features make the profile appear slightly doubled:
 cross-correlation of the ``star-spot" model spectra with the template
 G0 spectrum also shows that double structure.

	Clearly, the ``star-spot" model fits the observations of FU Ori
 much better than either disk model. Note that in making this fit, we consider
relatively weak photospheric lines of metals having EW $\approx$ 0.1 \AA. 
The structure of stronger lines often includes shortward-shifted absorption
features formed in the expanding shell and, in some cases, emission cores.
Figure 11 shows the observed spectrum of FU Ori around the 
\ion{Li}{1} $\lambda$6707 line on two occasions: the Li line is strong due
to the enhanced Li abundance in FU Ori; the same section calculated with
 the ``star-spot" model is shown for comparison. The two
 spectra of FU Ori differ in the shortward wing: the deep absorption at
--40 km\,s$^{-1}$ is prominent on 2005 November 23 (it 
is also present at lesser strength in other spectra of FU Ori) and, when
present, causes the line to appear double. Comparison to the ``star-spot"
 model shows that any emission core, if present, is very weak. Such
 shortward-shifted shell lines tend to be stronger in the blue region.

\section{Summary}

	There is no doubt that FU Ori possesses a disk: radiation from
the disk and envelope dominates the IR, the outline of the disk was
 resolved at AU scale by IR interferometry \citep{mal05}, and the
 distribution of energy in the infrared is compatible with a disk model. 
So our question was: what is it that is seen in the optical region, a
central star or an inner accretion disk?

	We found that
(1) all weak photospheric lines have the same line width and profile,
as expected for a rigidly rotating body, but in definite conflict with
prediction for the self-luminous Keplerian disk model;
(2) those profiles can be explained if the central object is a rapidly
rotating high-luminosity star with a dark polar spot; and
(3) there is no sign of the line doubling, or the dependence of line width
on wavelength (in the optical region) that is expected for the disk
model.

	In any disk, the temperature and rotational velocity change with
radius, and this defines the width of spectral lines in integrated light.
It would be strange if the disk temperature and velocity profiles were
designed by nature in such a way that all atomic lines formed in different
parts of the disk were of equal width.  It is more natural to assume that
the optical spectrum of FU Ori is produced at the central object, while
the IR spectrum is formed in the circumstellar disk (as in classical T Tauri stars),
and the 2.3 $\mu$m CO lines---a characteristic feature of FUors---are
formed in a cool disk or a distant shell. Certainly there is no evidence
of TiO bands or other M-type features  in the optical region: see Figure 9.

	Do these results undermine the self-luminous accretion disk
hypothesis in a significant way? No, but they do demonstrate that some
modification is in order.  It is remarkable that the hypothesis still stands
even though two of the strongest observational arguments
that were originally urged in its favor are now seen, in the case of the
prototype, to be invalid.

\acknowledgements 
 P. P. acknowledges the support of grant INTAS 03-516311 during this
investigation, and G. H. is indebted to the National Science Foundation,
grant AST 02-04021 for partial support.  We thank Ann Boesgaard, who obtained
the two 2002 HIRES spectrograms of FU Ori at our request, and Suzan
Edwards, Bo Reipurth, and Steve Stahler for helpful comments.

\clearpage

\begin{deluxetable}{ccccc}
\tablecolumns{5}
\tabletypesize{\small}
\tablewidth{0pc}
\tablecaption{ HWHD, Averaged over the Lines Listed in Table 2, 
 and $\Delta$RV, Radial Velocity Shifts by Cross-Correlation }
\tablehead{ \colhead{} & \colhead{}   & \colhead{HWHD} &
 \colhead{$\sigma$}  & \colhead{$\Delta$RV }  \\
 \colhead{Date } & \colhead{JD } & \colhead{(km s$^{-1}$)} &
 \colhead{(km s$^{-1}$)} & \colhead{(km s$^{-1}$})}  

\startdata

2003 Jan 11& 52650.85&  61.77&   0.29&  0.56 \\
2003 Feb 11& 52681.81&  61.64&   0.27&  0.10 \\
2004 Sep 24& 53273.09&  61.96&   0.29&  0.31 \\
2004 Nov 21& 53331.05&  61.77&   0.24& -0.31 \\
2005 Nov 23& 53698.12&  62.04&   0.28& -0.53 \\
2006 Dec 10& 54079.97&  62.16&   0.23&  0.29 \\
\enddata
\label{Table 1}
\end{deluxetable}

\begin{deluxetable}{cccc|cccc}
\tablecolumns{8}
\tablewidth{0pc}
\tabletypesize{\small}
\tablecaption{Lines Selected for Measurement of the HWHD at the
 Longward Wing}
\tablehead{
\colhead{$\lambda$} & \colhead{} & \colhead{EP}&
 \colhead{} & 
\colhead{$\lambda$} & \colhead{} & \colhead{EP}&
 \colhead{} \\
\colhead{(\AA)} & \colhead{Ion } & \colhead{(eV)} & \colhead{Note\tablenotemark{a}} & 
\colhead{(\AA)} & \colhead{Ion } & \colhead{(eV)} & \colhead{Note\tablenotemark{a}} }
\startdata
 5383.37 &  \ion{Fe}{1} &  4.31 & \nodata & 6726.66 &  \ion{Fe}{1} &  4.61 & r \\
 5717.83 &  \ion{Fe}{1}&  4.28 & \nodata & 6767.79 &  \ion{Ni}{1} &  1.83 & - \\
 5772.15 &  \ion{Si}{1}&  5.08& \nodata & 6810.26&  \ion{Fe}{1}&  4.61& -\\
 5775.08&  \ion{Fe}{1}&  4.22& \nodata & 6814.94&  \ion{Co}{1}&  1.96& r\\
 5862.35&  \ion{Fe}{1}&  4.55& \nodata & 6828.59&  \ion{Fe}{1}&  4.64& r\\
 5899.29&   \ion{Ti}{1}&  1.05& \nodata & 7090.38&  \ion{Fe}{1}&  4.23& r\\
 5922.11&   \ion{Ti}{1}&  1.05& \nodata & 7122.19&  \ion{Ni}{1}&  3.54& -\\
 5934.66&  \ion{Fe}{1}&  3.93& \nodata & 7344.70&   \ion{Ti}{1}&  1.46& r\\
 5965.83&   \ion{Ti}{1}&  1.88& \nodata & 7393.60&  \ion{Ni}{1}&  3.61& -\\
 5987.07&  \ion{Fe}{1}&  4.80& r & 7445.75&  \ion{Fe}{1}&  4.26& r\\
 6016.66&  \ion{Fe}{1}&  3.55& \nodata & 7511.02&  \ion{Fe}{1}&  4.18& -\\
 6024.06&  \ion{Fe}{1}&  4.55& r & 7525.11&  \ion{Ni}{1}&  3.63& r\\
 6027.05&  \ion{Fe}{1}&  4.09& \nodata & 7555.60&  \ion{Ni}{1}&  3.85& -\\
 6056.01&  \ion{Fe}{1}&  4.73& r & 7568.89&  \ion{Fe}{1}&  4.28& -\\
 6108.11&  \ion{Ni}{1}&  1.68& \nodata & 7574.04&  \ion{Ni}{1}&  3.83& -\\
 6180.20&  \ion{Fe}{1}&  2.73& \nodata & 7586.01&  \ion{Fe}{1}&  4.31& r\\
 6270.23&  \ion{Fe}{1}&  2.86& \nodata & 7727.61&  \ion{Ni}{1}&  3.68& r\\
 6355.03&  \ion{Fe}{1}&  2.85& r & 7780.55&  \ion{Fe}{1}&  4.47& -\\
 6358.70&  \ion{Fe}{1}&  0.86& \nodata & 7788.94&  \ion{Ni}{1}&  1.95& -\\
 6380.74&  \ion{Fe}{1}&  4.19& r & 7855.44&  \ion{Fe}{1}&  5.06& -\\
 6411.65&  \ion{Fe}{1}&  3.65& \nodata & 7912.87&  \ion{Fe}{1}&  0.86& -\\
 6439.08&   \ion{Ca}{1}&  2.53& \nodata & 7937.13&  \ion{Fe}{1}&  4.33& -\\
 6471.66&   \ion{Ca}{1}&  2.53& r & 8075.15&  \ion{Fe}{1}&  0.92& r\\
 6475.62&  \ion{Fe}{1}&  2.56& r & 8080.55&  \ion{Fe}{1}&  3.30& -\\ 
 6569.22&  \ion{Fe}{1}&  4.73& \nodata & 8085.18&  \ion{Fe}{1}&  4.45& r\\ 
 6581.21&  \ion{Fe}{1}&  1.49& \nodata & 8426.51&   \ion{Ti}{1}&  0.83& r\\ 
 6586.31&  \ion{Ni}{1}&  1.95& \nodata & 8611.80&  \ion{Fe}{1}&  2.85& -\\
 6707.89&  \ion{Li}{1}&  0.00& \nodata & 8621.60&  \ion{Fe}{1}&  2.95& -\\
 6717.68&   \ion{Ca}{1}&  2.71& r & 8648.47&   \ion{Si}{1}&  6.21& -\\
 6721.85&  \ion{Si}{1}&  5.86& \nodata &        &     &      &  \\
\enddata
\tablenotetext{a}{$r$ = only the longward half of that profile is
 blend-free.}

\label{Table 2}
\end{deluxetable}

 \begin{deluxetable}{cccc|cccc}
\tablecolumns{8}
\tablewidth{0pc}
\tablecaption{HWHD of Selected Lines in the Composite Spectrum of
 Accretion Disk}
\tabletypesize{\small}
\tablehead{
 \colhead{$\lambda$} & \colhead{} & \colhead{EP} & \colhead{ HWHD} &
 \colhead{$\lambda$} & \colhead{} & \colhead{EP} & \colhead{ HWHD} \\
 \colhead{(\AA)} & \colhead{Ion} & \colhead{(eV)} & \colhead{(km s$^{-1}$)} &
 \colhead{(\AA)} & \colhead{Ion} & \colhead{(eV)} & \colhead{(km s$^{-1}$)} 
 } 

 \startdata

5965.83 & \ion{Ti}{1} & 1.88 &  49.9 & 7525.11 & \ion{Ni}{1} &  3.63 &  54.2 \\
6016.66 & \ion{Fe}{1} & 3.55 &  66.8 & 7555.59 & \ion{Ni}{1} &  3.85 &  57.6 \\
6024.06&  \ion{Fe}{1} & 4.55 &  59.3 & 7568.89 & \ion{Fe}{1} &  4.28&  53.7\\ 
6027.05& \ion{Fe}{1}  & 4.08 &  53.7 & 7574.04 & \ion{Ni}{1} &  3.83&  55.6\\
6039.72& \ion{V}{1}   &  1.06&  47.1& 7586.01  & \ion{Fe}{1} &  4.31&  56.8\\
6358.69& \ion{Fe}{1}  &  0.86&  52.5& 7714.31  & \ion{Ni}{1} &  1.93&  52.3\\ 
6371.37& \ion{Si}{2}  &  8.12&  74.8& 7727.61  & \ion{Ni}{1} &  3.68&  58.5\\
6569.22& \ion{Fe}{1}  &  4.73&  60.0& 7738.96  & \ion{Ti}{1} &  1.46&  41.9\\
6581.21& \ion{Fe}{1}  &  1.49&  47.4& 7767.44  & \ion{Ti}{1} &  0.90&  40.0\\
6586.31& \ion{Ni}{1}  &  1.95&  54.7& 7788.94  & \ion{Ni}{1} &  1.95&  49.8\\
6707.89& \ion{Li}{1}  &  0.00&  50.5& 7912.87  & \ion{Fe}{1} &  0.86&  45.8\\ 
6717.68& \ion{Ca}{1}  &  2.71&  59.3& 7937.13  & \ion{Fe}{1} &  4.33&  57.0\\
6721.85& \ion{Si}{1}  &  5.86&  69.3& 8075.15  & \ion{Fe}{1} &  0.92&  47.9\\
6726.66& \ion{Fe}{1}  &  4.61&  57.5& 8080.54  & \ion{Fe}{1} &  3.30&  50.3\\
6767.77& \ion{Ni}{1}  &  1.83&  57.2& 8085.18  & \ion{Fe}{1} &  4.45&  57.3\\ 
6772.31& \ion{Ni}{1}  &  3.66&  61.4& 8611.80  & \ion{Fe}{1} &  2.85&  48.6\\
6806.85& \ion{Fe}{1}  &  2.73&  52.2& 8621.60  & \ion{Fe}{1} &  2.95&  51.8\\
6814.94& \ion{Co}{1}  &  1.96&  48.8& 8648.46  & \ion{Si}{1} &  6.21&  59.5\\
7511.02& \ion{Fe}{1}  &  4.18&  57.1&          &             &      &      \\ 

\enddata
\label{Table 3}
\end{deluxetable}

\clearpage

\begin{figure}
 \includegraphics[height=3.0in,keepaspectratio=true,origin=c,angle=0.]
{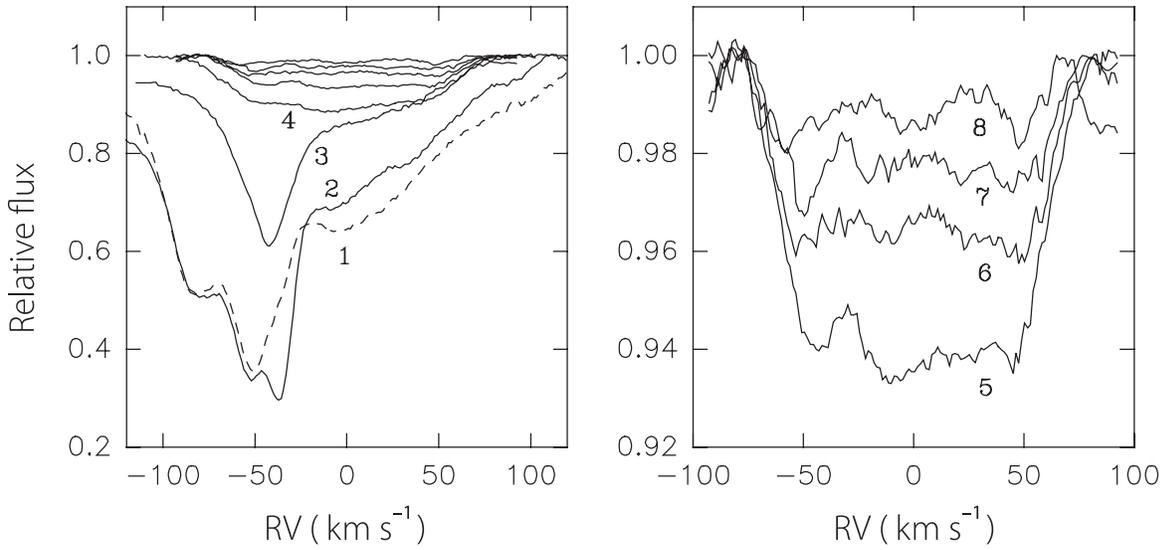}
\caption{Line profiles in FU Ori. Left panel: (1) \ion{Mg}{1} 5183.60, (2) \ion{Fe}{2} 5018.44,
 (3) \ion{Fe}{2} 5316.61, (4)  \ion{Ca}{1} 6439.08. Profiles of some weaker lines are expanded in
 the right panel:
(5) \ion{Fe}{1} 7511.02, (6) \ion{Ni}{1} 7393.60, (7) \ion{Fe}{1} 8080.55, (8) \ion{Fe}{1} 7107.46.
 No smoothing or filtering was performed on these spectra.  The S/N ratio
at the continuum level in the 6000--7000 \AA\ region is about 600. }
\label{Fig. 1} 
\end{figure}

\clearpage

\begin{figure}
 \includegraphics[height=3.0in,keepaspectratio=true,origin=c,angle=0.]
 {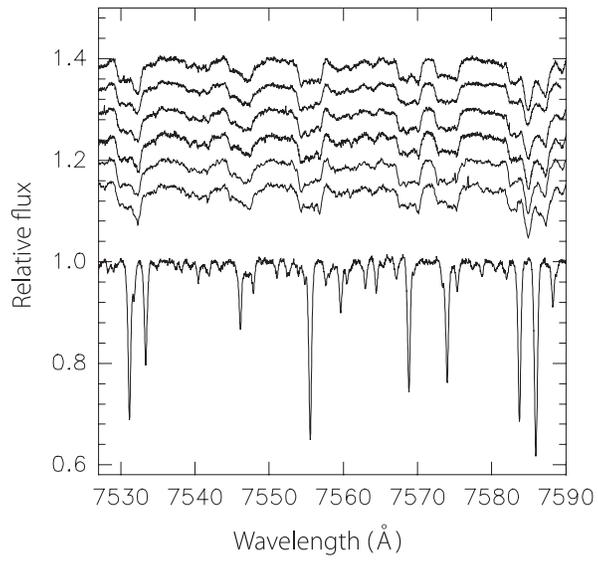}
\caption{Fragments of FU~Ori spectra of 2003--2006. The lower spectrum is of
$\beta$ Aqr (G0 Ib).}
\label{Fig. 2} 
\end{figure}

\clearpage

\begin{figure}
\includegraphics{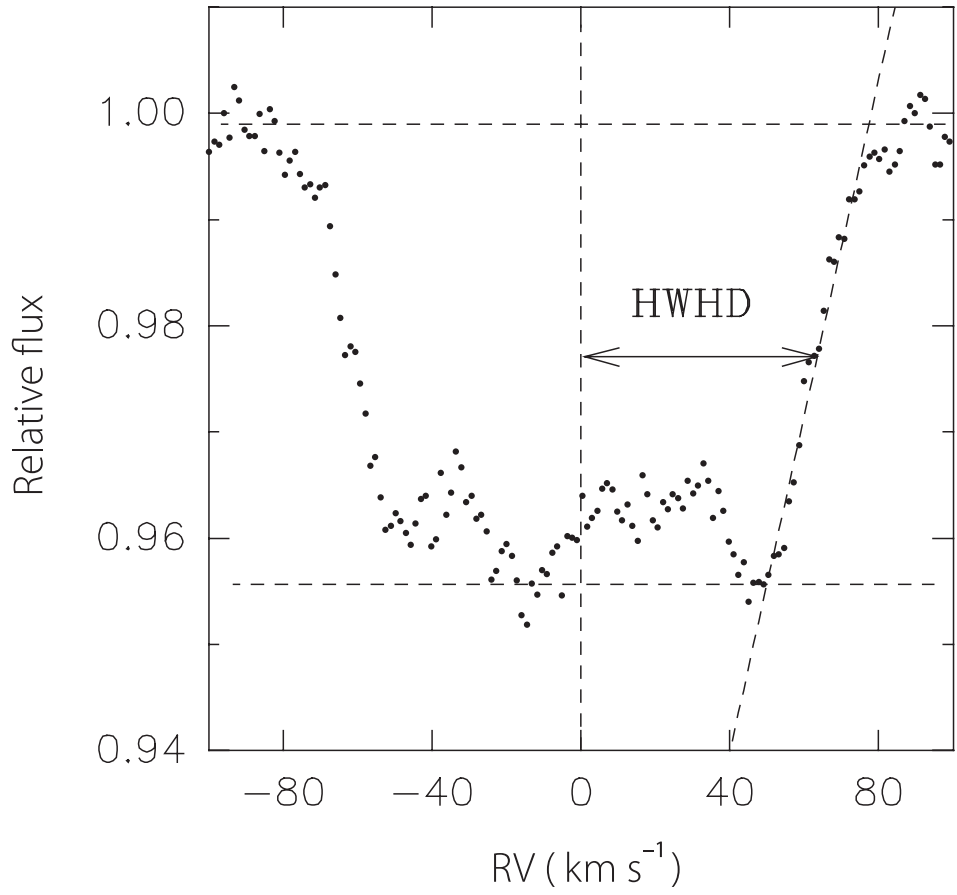}
\caption{Line profile of \ion{Fe}{1} 7568.89 and definition of the ``half-width
 at half-depth" (HWHD).}
\label{Fig. 3} 
\end{figure}

\clearpage

\begin{figure}
\includegraphics[height=3.0in,keepaspectratio=true,origin=c]{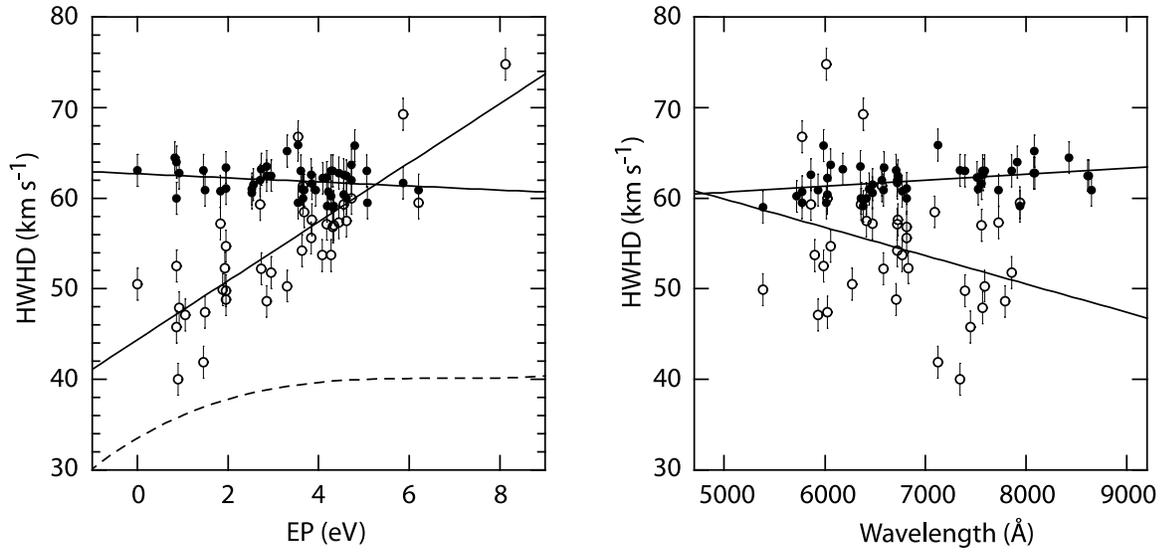}
\caption{ The dependence of line width (HWHD) on excitation potential (EP)
 is shown in the left panel, and its dependence on wavelength in the right.
 Filled circles indicate observed values, open circles those predicted by the
model.  The significance of the regression line slopes is discussed
in the text.  The data are from the spectrum of 2004 November 21. }
\label{Fig. 4} 
\end{figure}

\clearpage

\begin{figure}
\includegraphics[height=3.5in,keepaspectratio=true,origin=c]{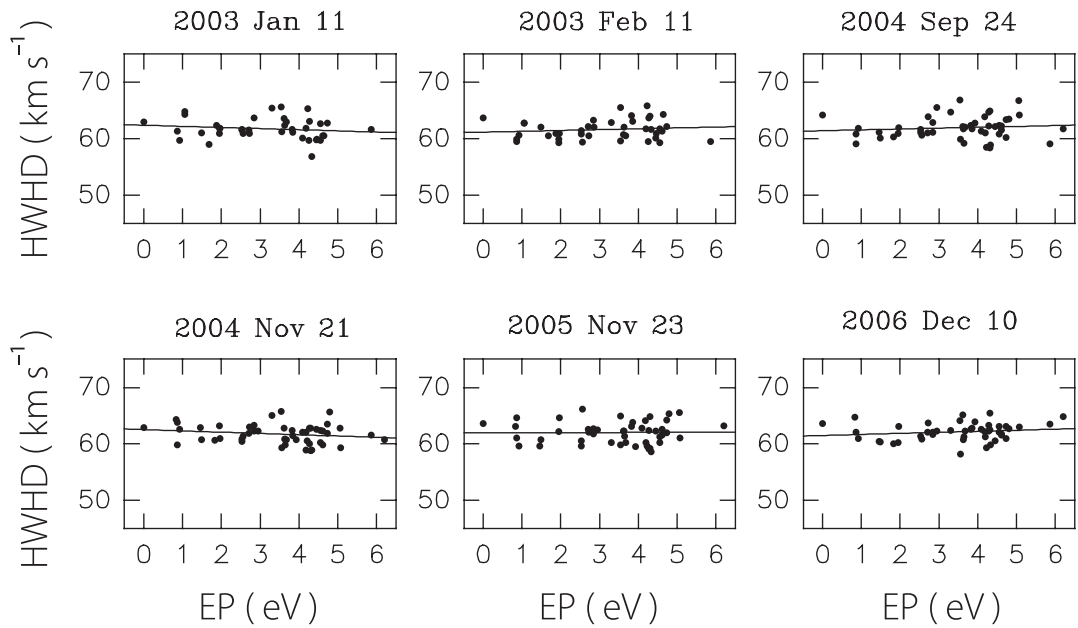}
\caption{HWHD versus EP in each of the six HIRES spectra of FU Ori.}
\label{Fig. 5} 
\end{figure}

\clearpage

\begin{figure}
\includegraphics[height=2.5in,keepaspectratio=true,origin=c]{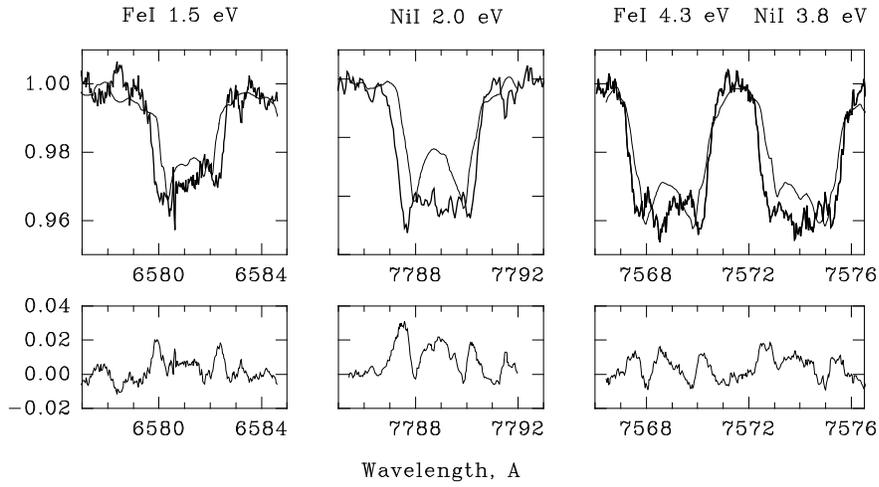}
\caption{Comparison of observed line profiles (thick) to those calculated
 in the disk model (thin). Note that the width of the modeled profiles
 depends on EP.  In the first and third panels, \ion{Fe}{1} 6581.21,
\ion{Fe}{1} 7568.89 and \ion{Ni}{1} 7574.04 were taken from the exposure
of 2004 November 21; in the second panel, \ion{Ni}{1} 7588.94 is from 2003
February 11.  The bottom panels show the differences, model minus observed,
in units of continuum intensity. }
\label{Fig. 6}
\end{figure}

\clearpage

\begin{figure}
\includegraphics[height=3.0in,keepaspectratio=true,origin=c]{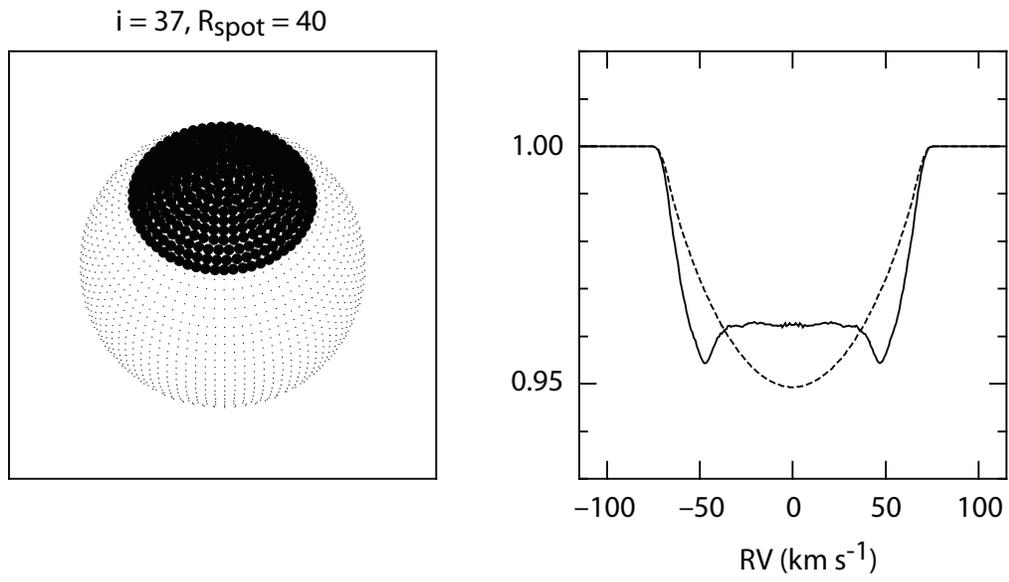}
\caption{ Left panel: model of a star with a dark polar spot.
 Right: typical line profile in the star with the polar spot (solid) as
compared with unspotted star (dashed).}
\label{Fig. 7}
\end{figure}

\clearpage

\begin{figure}
 \includegraphics[height=3.0in,keepaspectratio=true,origin=c]{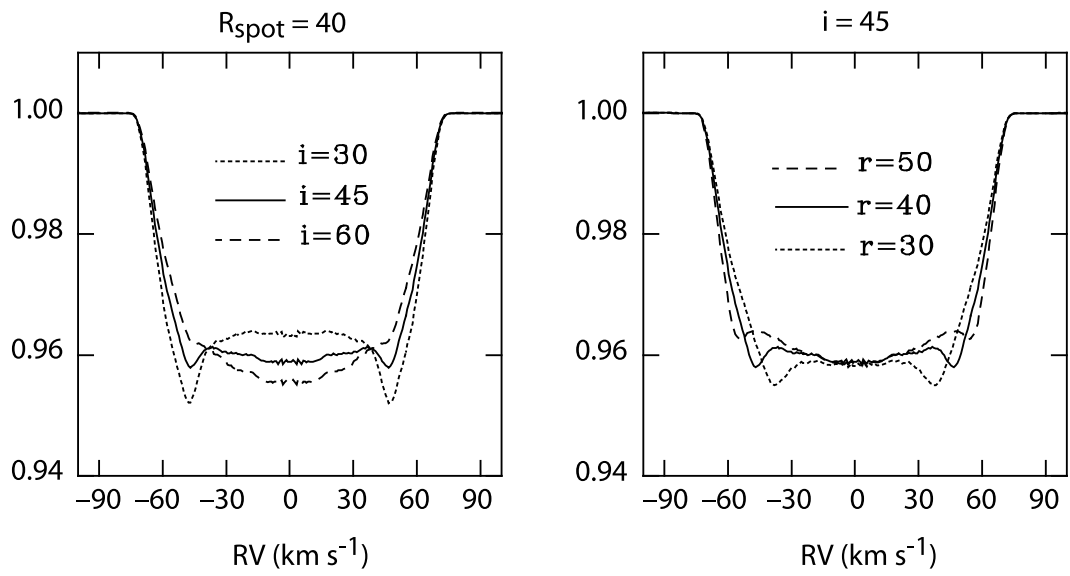}
\caption{ Line profiles in the ``star-spot" model as a function of spot radius
and inclination of the rotational axis to line of sight.  Spot radius and
inclination are in degrees. }
\label{Fig. 8}
\end{figure}

\clearpage
\begin{figure}
\includegraphics[height=3.0in,keepaspectratio=true,origin=c]{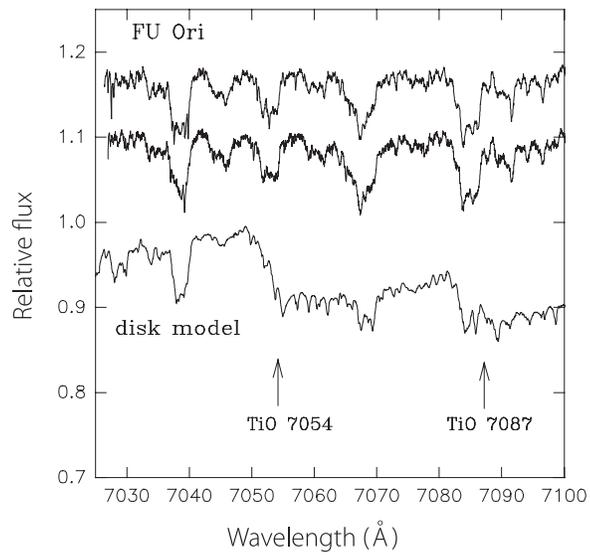}
\caption{The region of the $\lambda$ 7054 TiO bands in observed spectra of
  FU Ori (upper: 2004 September 24, lower: 2006 December 10) and in the
 disk model.}
\label{Fig. 9}
\end{figure}

\clearpage

\begin{figure}
\includegraphics[height=2.5in,keepaspectratio=true,origin=c]{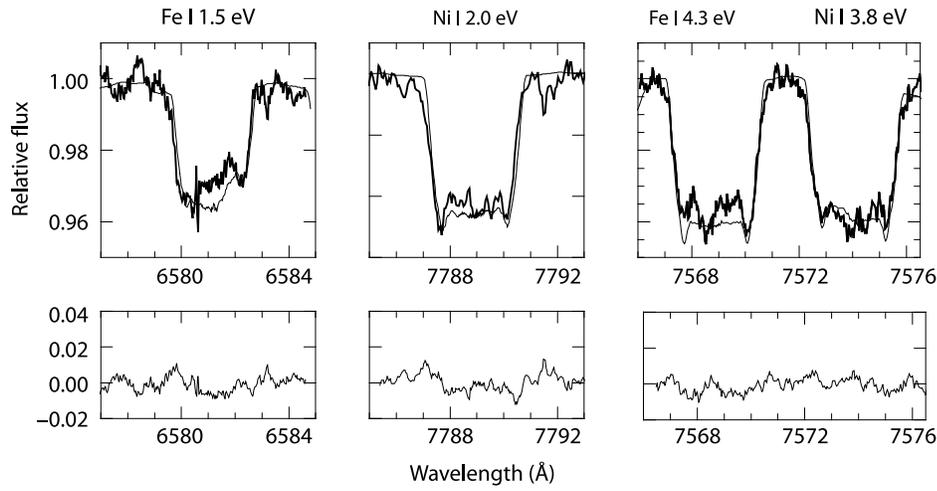}
\caption{Comparison of observed line profiles (thick) to those calculated
 in the ``star-spot" model (thin). The lines are the same as in Fig. 6.  As
in Fig. 6, the bottom panels show the differences, model minus observed,
in units of continuum intensity. }

\label{Fig. 10}
\end{figure}

\clearpage

\begin{figure}
\includegraphics[height=3.0in,keepaspectratio=true,origin=c]{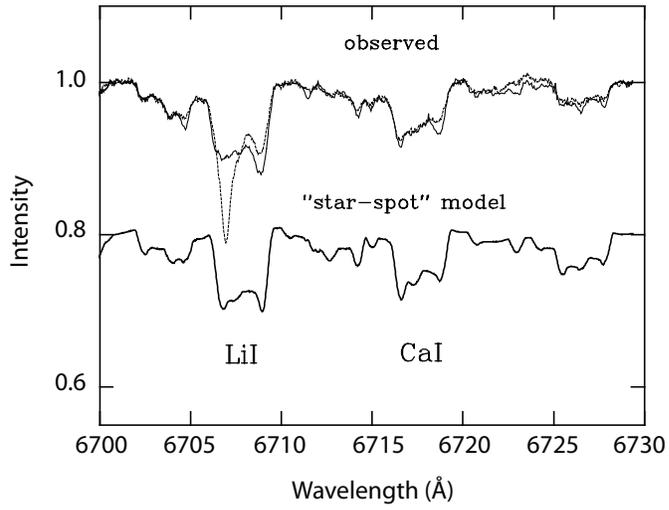}
\caption{Top: the region of \ion{Li}{1} $\lambda$ 6707 in FU Ori on 2003
 January 11 (heavy solid line) and on 2005 November 23 (light line). 
 Bottom: the same region from the ``star-spot" model. }
\label{Fig. 11}
\end{figure}


\end{document}